\documentclass[reprint, amsmath,amssymb, aps, prl]{revtex4-1}
\usepackage{graphicx}

\begin{document}
\title{Structure and scaling laws of liquid/vapor interfaces close to the critical point}

\author{Gy\"orgy Hantal}
\affiliation{Institute of Physics and Materials Science, University of Natural Resources and Life Sciences, Peter Jordan Straße 82, A-1190 Vienna, Austria}
\author{P\'al Jedlovszky}%
\affiliation{%
Department of Chemistry, Eszterh\'azy K\'aroly University, Le\'anyka utca 6, H-3300 Eger, Hungary
}%
\author{Marcello Sega}
\email{m.sega@fz-juelich.de}
\affiliation{
Forschungszentrum J\"ulich GmbH, Helmholtz Institute Erlangen-N\"urnberg for Renewable Energy (IEK-11), Cauerstr.\,1, D-91058 Erlangen, Germany}

\date{\today }

\begin{abstract}
\noindent{} To reach a deeper understanding of fluid interfaces it is necessary to identify a meaningful coarse-graining length that separates intrinsic fluctuations from  capillary ones, given the lack of a proper statistical mechanical definition of the latter. Here, with the help of unsupervised learning techniques, we introduce a new length scale based on the local density of the fluid. This length scale follows a scaling law that diverges more mildly than the bulk correlation length upon approaching the critical point. This allows to distinguish regimes of correlated and uncorrelated capillary waves from that of intrinsic fluctuations.
\end{abstract}

\maketitle

\noindent{}The detailed understanding of liquid/vapor interfaces  has proven to be an elusive objective, because of the interplay between fluctuations in the bulk and surface capillary excitations. The effect of bulk fluctuations on the structure of the interface is captured by theories such as that of  van~der~Waals\cite{van_der_waals_thermodynamische_1893,rowlinson_molecular_2002}, whereas Buff, Lovett, and Stillinger\cite{buff_interfacial_1965} were the first to place capillary fluctuations in the focus. Bulk phase fluctuations are expected to become dominant and cause the divergence of the interface width close to the critical point, while far from it the intrinsic interface thickness should be of the order of the molecular size for simple liquids and dominated by entropic fluctuations. Several approaches tried to reconcile these two pictures, starting with that of Weeks\cite{weeks_structure_1977},  who considered the statistical mechanics of columns of fluid separated by the bulk correlation distance \(\xi\). Several alternative routes have been made since then to overcome the limitations of the capillary wave theory, for example including a wave-vector dependency into the continuous Hamiltonian that describes the coarse-grained dynamics of the interface\cite{mecke_effective_1999,parry_pair_2014,hofling_enhanced_2015}, exploring nonlocal models\cite{parry_nonlocality_2004,fernandez_intrinsic_2013,hernandez-munoz_capillary_2016} or, recently, extending density functional models to asymmetric interfaces\cite{parry_local_2016,parry_goldstone_2019}. The concept of columnar averaging has been applied with success, for example, in simulations of polymer mixtures\cite{werner_anomalous_1997,werner_intrinsic_1999}. Other computational strategies\cite{chacon_intrinsic_2003, jorge_intrinsic_2007,partay_new_2008, willard_instantaneous_2010,sega_generalized_2013} have taken alternative routes to the definition of the liquid/vapor interface and are typically not involving columnar averages for the calculation of intrinsic profiles because the definition of the interface position is provided at atomistic resolution. However, all these approaches are characterised by the presence of a free parameter that directly or indirectly sets the separation between capillary fluctuations and corrugations happening below that scale (intrinsic fluctuations).

No theory or computational approach has provided so far a comprehensive and unambiguous framework to describe fluid interfaces. A particularly contentious topic is the choice of the bulk correlation length \(\xi\) as a coarse-graining scale. This choice has the obvious advantage of separating the surface into uncorrelated regions, making the problem amenable to analytical treatment, seemingly solving the problem of providing a definition for the interface location. However, there is no fundamental need to make this choice. In fact, as we will show, it is possible to define an alternative coarse-graining length, based on the local density, to separate capillary from intrinsic fluctuations, which provides further insight into the properties of capillary waves and the structure of liquid interfaces.

The condition of zero surface excess  
\begin{equation}
\int_{-\infty}^{z_{eq}} \rho_L -\rho(z) dz= \int_{z_{eq}}^{\infty}\rho(z) -  \rho_V dz,
\label{eq:equimolar}
\end{equation}
provides an implicit definition for the location of the equimolar surface plane \(z=z_{eq}\) along the surface normal \(z\) for a single component system with density profile \(\rho(z)\) and bulk values \(\rho_L\) and \(\rho_V\) in the liquid and  vapor regions, respectively. For a periodic simulation cell of edge \((0,L)\) along \(z\) with the liquid phase appearing as a slab in the middle of the box, this is simply given by
\begin{equation}
        z_{eq} = \frac{L}{2} + \frac{\overline\rho-\rho_V}{\rho_L-\rho_V} \frac{ L}{2}.\label{eq:zeq}
\end{equation}
\begin{figure*}
\includegraphics[width=\textwidth]{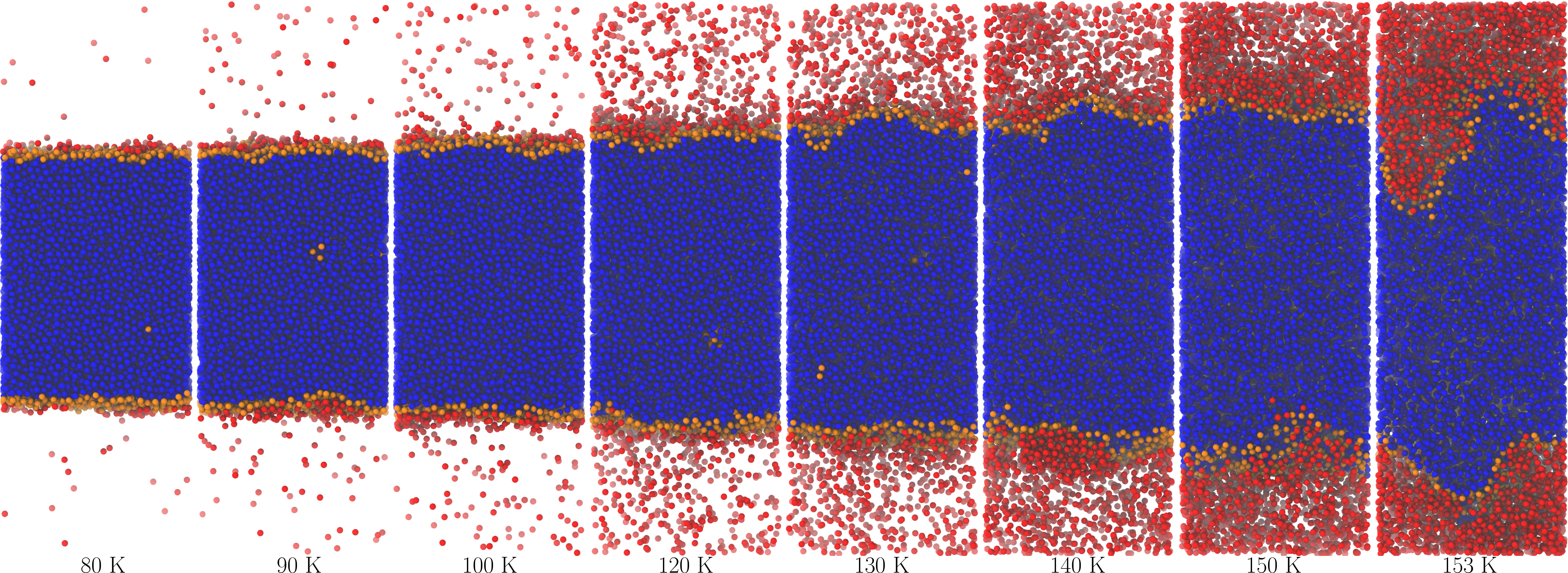}
\caption{Snapshots of the argon system at different temperatures. The blue and orange colors  mark atoms belonging to the largest cluster of liquid-like molecules, with orange denoting the surface ones as determined using the optimal probe sphere radius (different for each temperature). The red color marks the remaining atoms, independently whether they belong to smaller liquid droplets or to low density, vapor-like regions.}
\label{fig:poly-Ar}
\end{figure*}
As Stillinger  noted\cite{stillinger_capillary_1982}, the boundary of the liquid phase (however defined) does not need to coincide with the equimolar Gibbs dividing surface. If the density profile \(\rho(z)\) is piecewise constant and equal to the two bulk densities in the respective phases, then \(z_{eq}\) is located precisely between the two. When \(\rho_L\gg\rho_V \) there should be no ambiguity in this statement.
If the bulk values and the distribution on the liquid side remain unchanged, but the vapor phase develops an accumulation of material at the interface  (as in a simple mean-field theory),  \(\overline\rho\) has to increase, consequently moving \(z_{eq}\) toward the vapor phase and away from the actual surface of the liquid  phase. Here, we pick an alternative route to distinguish liquid from vapor, by monitoring the density of the local molecular environment. Practically, this task can be accomplished in an efficient way using unsupervised learning techniques such as the DBSCAN algorithm\cite{ester_density-based_1996}. 
Once molecules are assigned to the liquid and vapor phases, the (macroscopic) location of the liquid phase separating plane \(z_L\) can be defined equivalently to Eq.~(\ref{eq:zeq}), with the liquid phase density profile \(\rho_L(z)\) replacing \(\rho(z)\). One can think of \(z_L\) as the average position of an enveloping, corrugated surface that, in case \(\rho_L(z)\) drops to zero for sufficiently large values of \(z\), always contains all liquid molecules. 
The link to capillary waves is provided through the assignment of surface molecules. In fact, if one performs the thought experiment of carving out a portion of material from a bulk liquid, the location of \(z_L\) must be shifted with respect to the interfacial atomic centres by half of the average molecular size
\( \overline{r} \simeq \left(4\pi\rho_L/3\right)^{-1/3} \)
to yield correctly zero excess surface density. 
We use this constraint to determine the optimal probe sphere radius for the surface atom recognition algorithm GITIM\cite{sega_generalized_2013,sega_pytim_2018} by requiring the average position of the surface atoms \(z_S(R_p)\), parameterised by the probe sphere radius \(R_p\),  to satisfy \(z_S(R_\mathrm{cg}) + \overline{r} = z_L\), where the optimal probe sphere radius identifies the coarse-graining length scale  \(R_\mathrm{cg}\).

We tested these concepts on a series of molecular dynamics simulations of water\cite{abascal_general_2005} and argon\cite{rowley_monte_1975}, over a range of reduced temperature \(\tau = 1- T/T_c\) from about 0.5 to 0.03, where \(T_c\) indicates the critical temperature of the respective models. 
We performed molecular dynamics simulations of argon and water using the GROMACS simulation package, release 2018.2\cite{abraham_gromacs_2015}. We integrated the equations of motion in the canonical ensemble using the Verlet algorithm, 2~fs integration time step, Nos\'e-Hoover\cite{nose_molecular_1984,hoover_canonical_1985} thermostat with 1~ps time constant and constraining the geometry of water molecules \cite{miyamoto_settle_1992}. We computed the long range part of Coulomb and dispersion interactions using a particle-mesh Ewald method\cite{essmann_smooth_1995} with a grid spacing of 1.2~\AA, fourth order polynomial interpolation scheme, short-range cutoff of 12~\AA, metallic boundary conditions and a relative interaction strength at the cutoff of \(10^{-5}\) and \(10^{-3}\) for the Coulomb and dispersion terms, respectively. 
We simulated 20000 argon atoms and 13824 water molecules in rectangular unit cells of size 88\(\times\)88\(\times\)260 and 75\(\times\)75\(\times\)250~\AA\(^3\), respectively. We started from a slab configuration with normal along the \(z\) axis and integrated the equations of motions for 10 ns, the last 5 of which we used for production. Except for the lowest temperatures, each simulation started from the last configuration of the lower temperature run. We stored configurations to disk every ps for subsequent analysis using the MDAnalysis\cite{michaudagrawal_mdanalysis_2011} and Pytim\cite{sega_pytim_2018} analysis packages. We assigned molecules to the liquid or vapor phases according to the protocol reported first in our study on mixtures with high partial miscibility\cite{sega_phase_2017}. We define the liquid phase as the largest, connected cluster of molecules having a high-density local environment. With this choice, any smaller droplet is by definition assigned to the vapor phase.
We identified each system's surface molecules in each frame by using a series of different probe sphere radii \(R_p\). We collected the histograms of their position, \(\rho_S(z ;  R_p)\), as well as the histograms of the liquid-like and vapor-like molecules, \(\rho_L(z)\) and \(\rho_V(z)\), respectively.  The function  \(x_L - z_S(R_p) \) turned out to be always a monotonously decreasing function of \(R_p\). Eventually, we determine the zero of the function \(x_L - z_S(R_p) \) by linear interpolation between the two values closest (but with opposites sign) to zero.

\begin{figure}
    \centering
    {\includegraphics[trim=15 105 10 105 ,clip,width=\columnwidth]{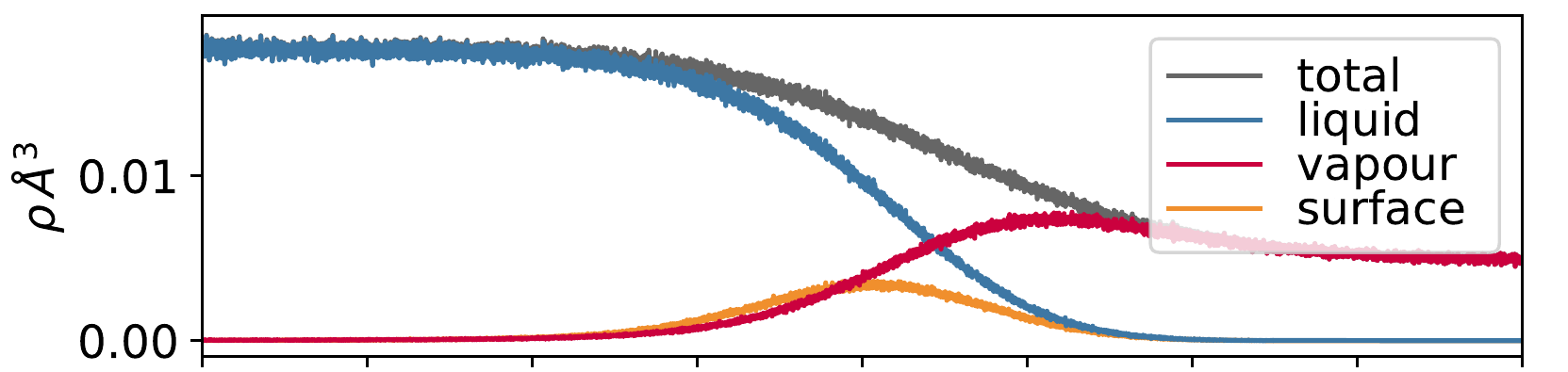}}
    {\includegraphics[trim=15 30  10 65 ,clip,width=\columnwidth]{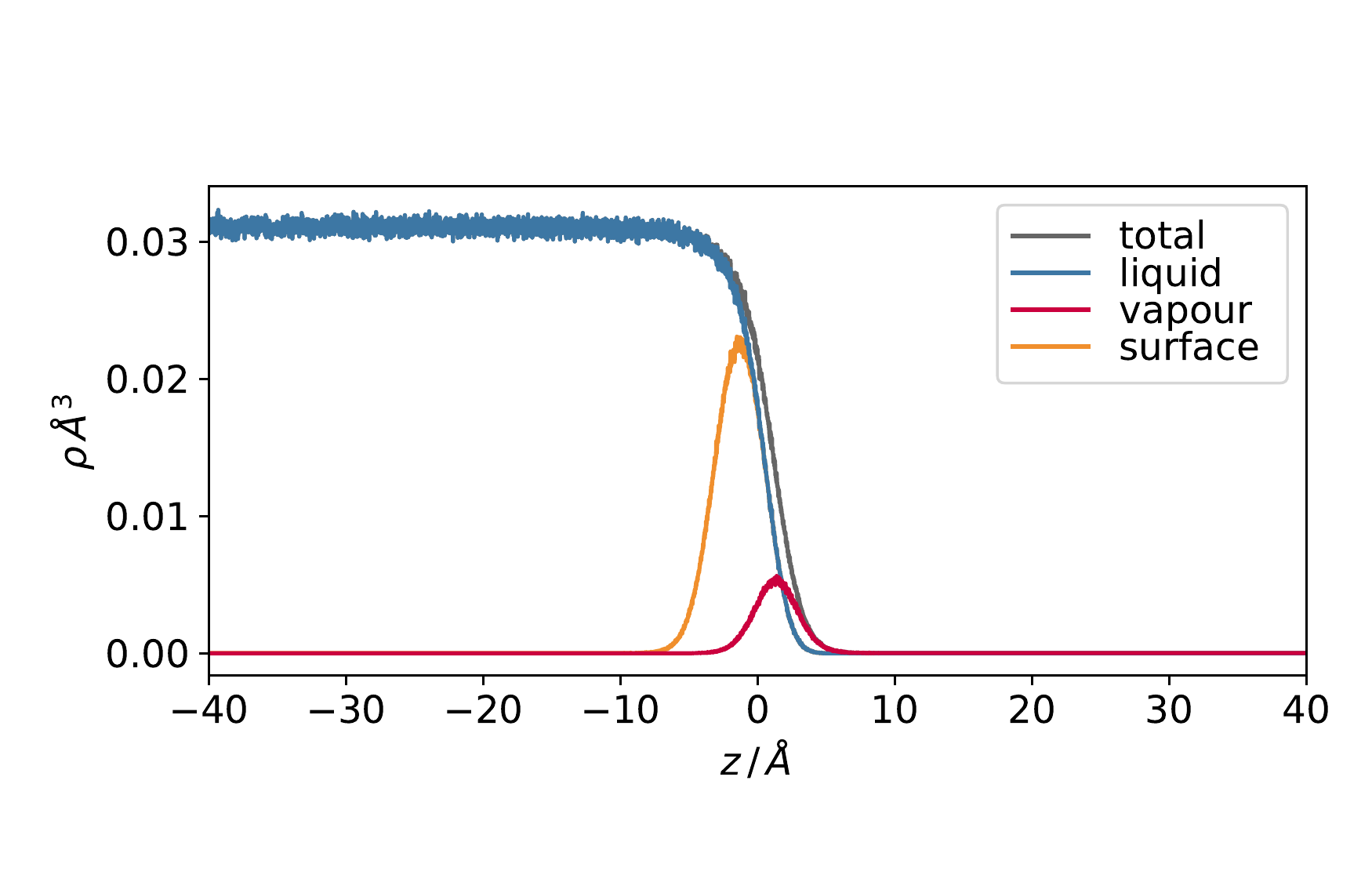}}
    \caption{Density profiles for argon at 150 (top) and 70 K (bottom). The total (gray), liquid phase (blue), vapor phase (red) and surface atoms (yellow) are shown, centered at \(z_L\).}
    \label{fig:profiles}
\end{figure}

Fig.~\ref{fig:poly-Ar} shows some snapshots of the argon system for a selected set of temperatures. There, atoms in the liquid phase are shown in blue (the internal ones) or orange (the surface ones), whereas atoms in the vapor phase are shown in red. The vapor phase, by design, includes also eventual liquid droplets disconnected from the main liquid phase. One of these droplets is evident in the \(T=140\)~K snapshot, where it is about to fuse with the liquid on the lower interface. The snapshots were produced using \(R_p =  R_\mathrm{cg}\). The interfacial atoms appearing in the bulk liquid phase mark spontaneous cavitation effects\cite{huang_cavity_2000}. 
In Fig.~\ref{fig:profiles} we report the density profiles of \(\rho(x)\), \(\rho_L(x)\), \(\rho_V(x)\), as well as the profile  \(\rho_S(x)\) of the surface atoms of the liquid phase for the argon liquid/vapor interface at \(T=150\) and 70~K, these two cases being representative of a liquid close and far from the critical point.
As expected, the vapor accumulates at the interface without becoming part of the liquid itself. We remind that according to the density-based clustering algorithm, all atoms which are not in a (liquid-like) high-density environment but still reachable from one of the liquid molecules would be considered as liquid ones. Atoms tagged as vapor ones are therefore clearly disconnected from the liquid phase in this sense.
\begin{figure}
    \centering
    \includegraphics[width=\columnwidth]{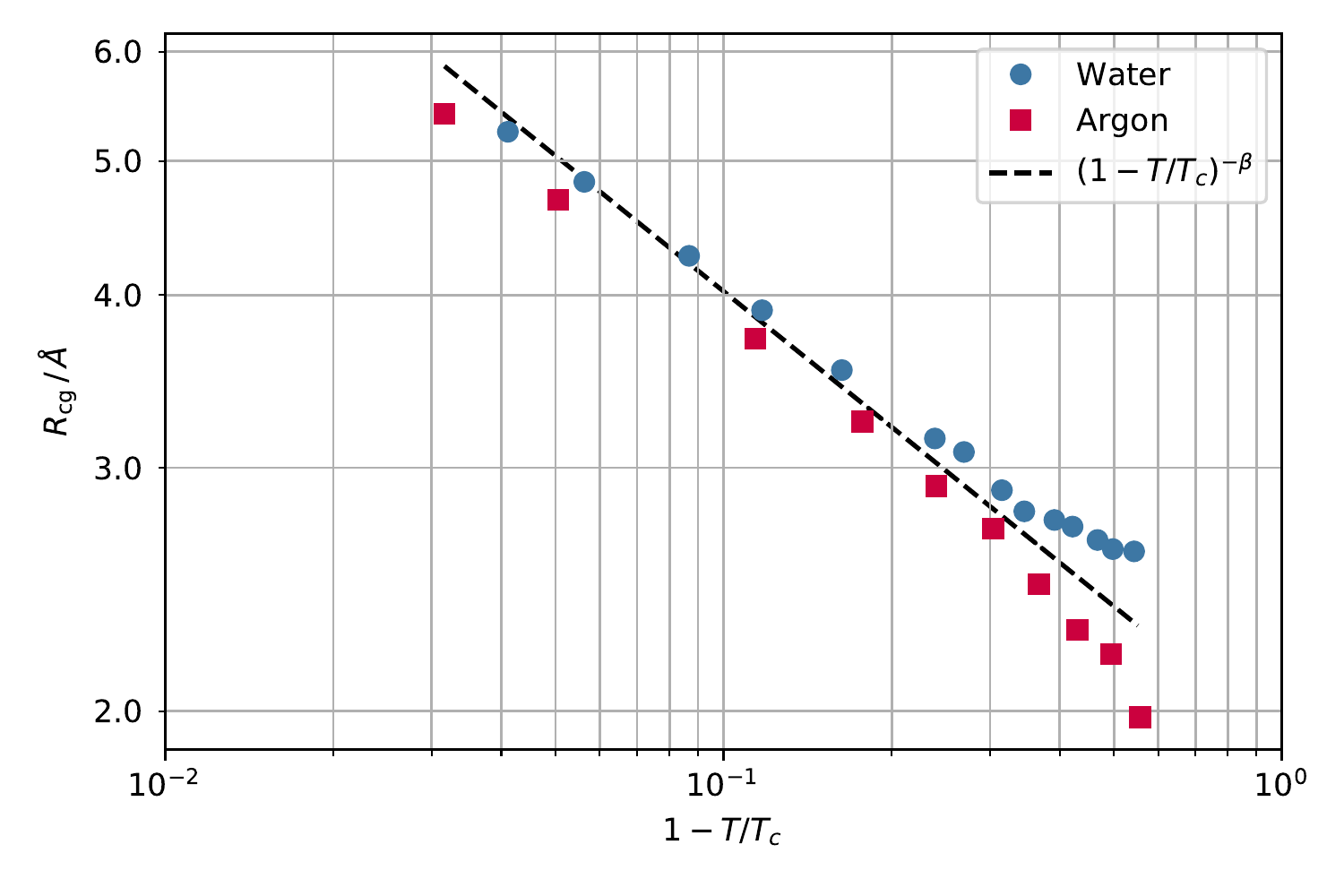}
    \caption{Coarse-graining length \(R_\mathrm{cg}\) as a function of the reduced temperature \(\tau = 1- T/T_c\) for water (circles) and argon (squares) as well as the scaling law \(\tau^{-\beta}\) (black dashed line).} \label{fig:Ropt}
\end{figure}

In Fig.~\ref{fig:Ropt} we report the values of the coarse-graining sphere radius as a function of the reduced temperature, which is the main result of this work. When approaching the critical temperature, the values of the coarse-graining radius show a universal behavior. This behavior is apparently compatible with \( R_\mathrm{cg} \sim \tau^{-\beta}\), where \(\beta\simeq0.326\) is  the exponent for the order parameter \( \rho_L - \rho_V \sim \tau^\beta\). 
The physical meaning of  \(R_\mathrm{cg}\) is clearly that of a coarse-graining length that separates intrinsic fluctuations of the interface from capillary waves. By choosing a probe sphere radius smaller (or larger) than \(R_\mathrm{cg}\), the enveloping surface defined by the surface atoms would be located inwards  (or outwards) with respect to the liquid separating one, \(z_L\). The appealing property of   \(R_\mathrm{cg}\) is that it depends implicitly only on the assignment of molecules to the liquid and vapor phases, and not on any other geometrical considerations.
Notably, the scaling is clearly not the same as the correlation length\cite{jasnow_critical_1984}  \(\xi\sim\tau^{-\nu}\), with \(\nu\simeq0.63\) (see Ref.~\onlinecite{chester_carving_2020} for recent theoretical estimates of critical exponents of the XY universality class). 
\begin{figure}
    \centering
    \includegraphics[width=\columnwidth]{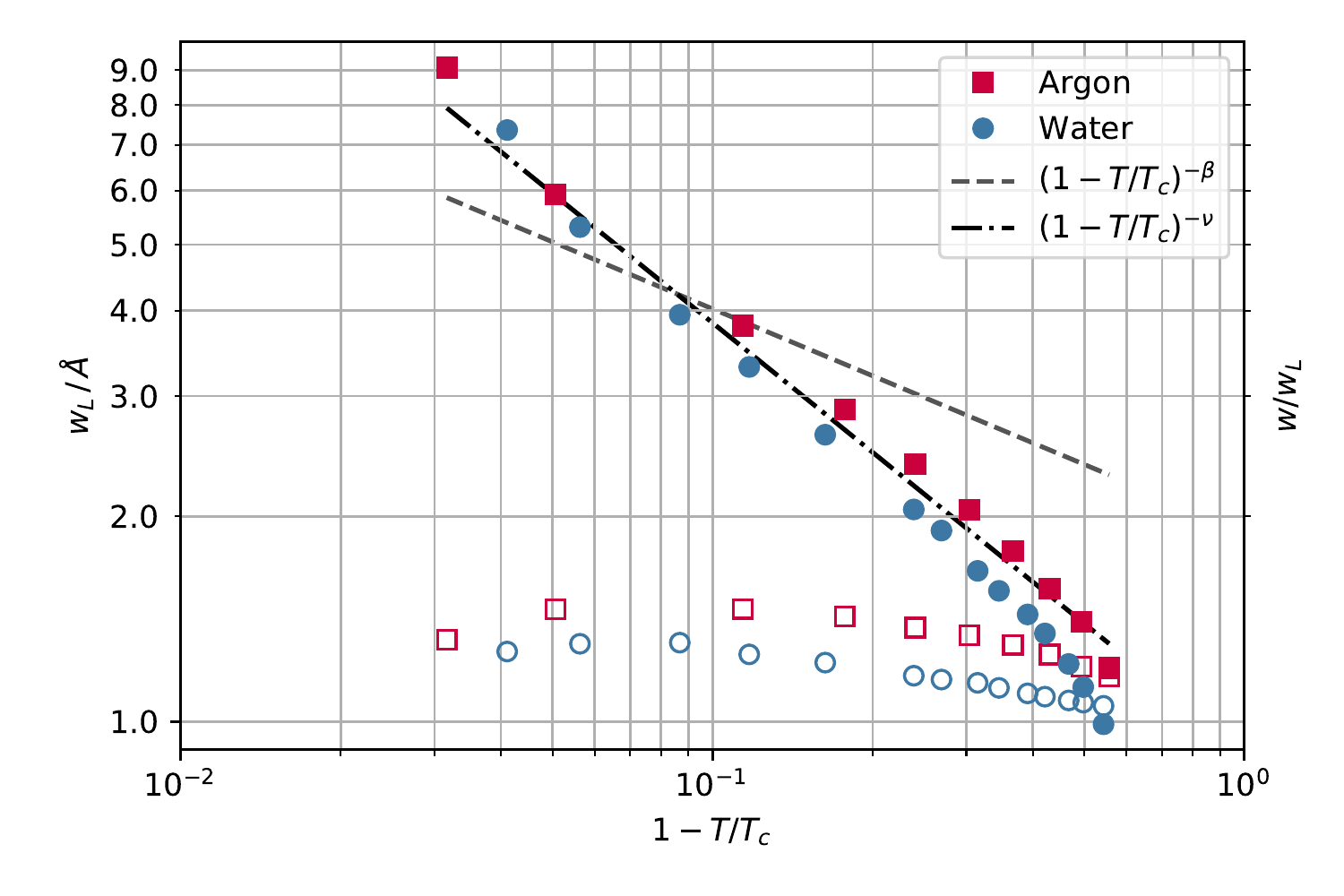}
        \caption{Apparent interfacial width \(w_L\) obtained from the profile \(\rho_L(z)\) for water (filled circles) and argon (filled squares). The ratio \(w/w_L\) between the the widths of \(\rho(z)\) and \(\rho_L(z)\) is also shown (open symbols). The scaling law \(\tau^{-\nu}\) is reported as a black dot-dashed line. For comparison, we report also the scaling law of \(R_\mathrm{cg}\)  as a light-gray, dashed line.}
    \label{fig:width}
\end{figure}
We computed also the apparent interfacial widths \(w\) and \(w_L\) by fitting the profiles \(\rho(z)\) and \(\rho_L(z)\) to a complementary error function, 
\(\rho(z) = \rho_V + \frac{1}{2}\rho_L \mathrm{erfc}\left(\frac{z-z_0}{2w}\right)\), the prediction of the convolution approximation\cite{jasnow_critical_1984} in the capillary wave  theory. We note in passing that the hyperbolic tangent solution of the Cahn-Hilliard theory\cite{cahn_free_1958} fits the profiles noticeably worse. 
The scaling of \(w_L\), which we show in Fig.~\ref{fig:width}, is compatible with the expected behavior \(\tau^{-\nu}\), with some deviation possibly appearing at high temperatures. The apparent width of the total profile \(w\) behaves similarly to, and is roughly a multiple of, \(w_L\). In Fig.~\ref{fig:width} we report also the scaling law previously found for \(R_\mathrm{cg}\). There is a crossing value for \(\tau\), \(\tau_x \simeq 0.1\), below which (i.e., at high temperature) the interface width is always larger than the coarse graining length \(R_\mathrm{cg}\). Since the former is representative of the bulk correlation length, this result shows that a part of the capillary waves spectrum at high temperature is necessarily composed of non-independent modes, even though they are, in the sense explained before, not intrinsic features of the interfaces. Of course, capillary wave theory would fail in this regime, as it assumes independent modes.
Conversely, it is possible (for \( \tau > \tau_x\)) that independent modes could be present even if they are intrinsic features of the interface. Notice that this does not exclude nonlocal effects, which are expected to become dominant below the scale of the bulk correlation length but can still be appreciated, for example for argon close (\(T\simeq 90\)~K) to the triple point, at distances in the order of nanometers\cite{fernandez_intrinsic_2013}.
\begin{figure}
    \centering
    \includegraphics[width=\columnwidth]{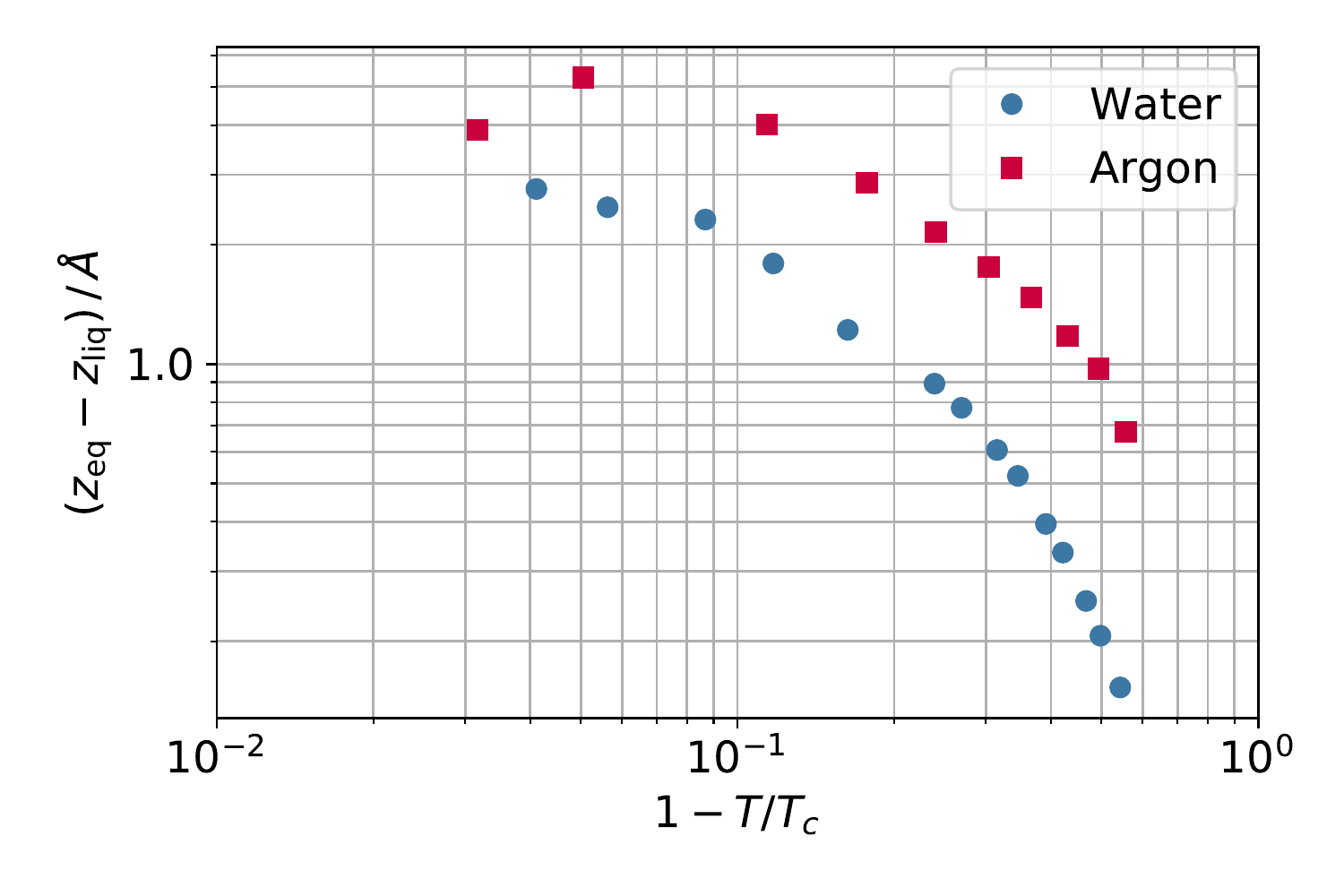}
        \caption{Separation between the equimolar (\(z_{eq}\)) and liquid (\(z_L\)) dividing surfaces as a function of the reduced temperature \(\tau = 1- T/T_c\) for water (circles) and argon (squares).}
    \label{fig:deltaZ}
\end{figure}
As mentioned in the beginning, the equimolar surface is not expected to coincide with the liquid separating surface even in simple ideal systems. We are now in a position to quantify this difference.  In fact, at low temperature, the separation between the two surfaces, \(z_{eq}-z_L\), is growing when the temperature increases, as reported in Fig.~\ref{fig:deltaZ}, for both argon and water.  However, this gap might not widen indefinitely when approaching the critical temperature, as the result for the highest simulated temperature for argon (\(T=153\)~K) seems to indicate.  

In conclusion, the procedure reported here provides a new coarse-graining length by requiring the average location of surface molecules to be compatible with that of the liquid separating plane. Strictly speaking, this is not a thermodynamic quantity like the Gibbs dividing surface because the set of molecules belonging to the liquid cannot be determined by thermodynamic means only. However, this internal self-consistency removes from a class of algorithms the dependence on arbitrary parameters, relying only on a recipe to separate liquid-like molecules from vapor-like ones. We are confident that this approach could be helpful to remove free parameters in other methods too. To our mind comes the instantaneous liquid interfaces method of Willard and Chandler\cite{willard_instantaneous_2010}, where one could choose the threshold density in such a way to match either the location of the liquid separating plane or the equimolar Gibbs dividing surface. Work in this direction is ongoing.

From a more general perspective, the presence of an optimal probe sphere size, \(R_\mathrm{cg}\), acquires a direct physical meaning because it sets the scale at which it makes sense to define the liquid surface. 
Below this coarse-graining length, fluctuations are not anymore living on the liquid/vapor interface but are penetrating into the liquid phase.  Since the scaling exponent of the present coarse-graining length scale is smaller than that of the bulk correlation one, part of these capillary waves are necessarily going to be correlated at a high enough temperature. Importantly, surface excitations with wavelengths between  \(R_\mathrm{cg}\) and \(\xi\), even though correlated,  should not be considered as intrinsic fluctuations of the liquid/vapor surface. 

\section*{Acknowledgements}
PJ acknowledges financial support from the NKFIH Foundation, Hungary (under project No. 134596). MS acknowledges support by the Deutsche Forschungsgemeinschaft (DFG) within the priority program SPP 2171, grant number 422794127.


%

\end{document}